\documentclass[fleqn,usenatbib]{mnras}

\usepackage{newtxtext,newtxmath}

\usepackage[T1]{fontenc}
\usepackage{ae,aecompl}

\usepackage{threeparttable}

\usepackage{graphicx}	
\usepackage{amsmath}	
\usepackage{amssymb}	

\title[Late-Time Observations]{Deep Late-Time Observations of the Supernova Impostors SN 1954J and SN 1961V}

\author[Patton, Kochanek, and Adams]{
Rachel A. Patton$^{1}$,
C. S. Kochanek$^{1,2}$,
and S. M. Adams$^{3}$
\\
$^{1}$Department of Astronomy, The Ohio State University, 140 West 18th Ave, Columbus, OH 43210, USA\\
$^{2}$Center for Cosmology and AstroParticle Physics, The Ohio State University,
191 West Woodruff Avenue, Columbus, OH 43210\\
$^{3}$Cahill Center for Astrophysics, California Institute of Technology, Pasadena, CA 91125, USA\\ 
}

\date{Accepted XXX. Received YYY; in original form ZZZ}

\pubyear{2018}

\begin{document}
\label{firstpage}
\pagerange{\pageref{firstpage}--\pageref{lastpage}}
\maketitle

\begin{abstract}
SN 1954J in NGC 2403 and SN 1961V in NGC 1058 were two luminous transients whose definitive classification as either non-terminal eruptions or supernovae remains elusive. A critical question is whether a surviving star can be significantly obscured by dust formed from material ejected during the transient. We use three lines of argument to show that the candidate surviving stars are not significantly optically extincted ($\tau \lesssim 1$) by dust formed in the transients. First, we use SED fits to new HST optical and near-IR photometry. Second, neither source is becoming brighter as required by absorption from an expanding shell of ejected material. Third, the ejecta masses implied by the H$\alpha$ luminosities are too low to produce significant dust absorption. The latter two arguments hold independent of the dust properties. The H$\alpha$ fluxes should also be declining with time as $t^{-3}$, and this seems not to be observed. As a result, it seems unlikely that recently formed dust can be responsible for the present faintness of the sources compared to their progenitors, although this can be verified with \textit{JWST}. This leaves three possibilities: 1) the survivors were misidentified; 2) they are intrinsically less luminous; 3) SN 1954J and SN 1961V were true supernovae.
\end{abstract}


\section{Introduction} \label{sec:intro}
There are a subset of massive stars that appear to experience violent eruptions, some of which are spectroscopically similar to Type IIn supernovae (SNe, \citeauthor{schlegel_new_1990} \citeyear{schlegel_new_1990}; \citeauthor{filippenko_optical_1997} \citeyear{filippenko_optical_1997}) due to their moderate line widths (v$_{ej} \lesssim$ 2000 km s$^{-1}$). The brightest of these eruptions are not easily distinguishable from the faintest SNe, leading to the term SN ``impostor" \citep{van_dyk_possible_2002} and the potential for misclassification. The only certain difference between the most luminous eruptions and the least luminous SNe is that these eruptions are non-terminal. These eruptions remain poorly understood and so accurate classification of these events is critical to understanding outburst mechanisms and rates. 

Two of the of the most famous examples are SN 1954J in NGC 2403 and SN 1961V in NGC 1058. SN 1954J was originally classified as an SN but was later identified as the luminous blue variable (LBV) V12 and reclassified as an eruption \citep{tammann_stellar_1968}. The progenitor remained fairly quiescent at M$_B \approx -$6.6 mag until it became highly variable a few years prior to its eruption, sometimes fluctuating by as much as two magnitudes over a few days. The peak of the transient was not observed due to its proximity to the Sun. When next observed, its magnitude had increased from M$_B$ $\approx -$8.5 mag to M$_B$ $\approx -$11.3 mag before settling at M$_B \approx -$5.6 mag, a full magnitude fainter than the progenitor \citep{tammann_stellar_1968}. The survivor never regained its pre-eruption luminosity and has not significantly varied since the eruption \citep{smith_post-eruption_2001, van_dyk_supernova_2005, kochanek_unmasking_2012, humphreys_tale_2017}. \citet{van_dyk_supernova_2005} resolved the region around SN 1954J into four stars (see Fig. \ref{fig:field}) and identified star 4 as the most likely survivor candidate due to its strong H$\alpha$ emission. Follow-up spectroscopy by \citet{humphreys_tale_2017} shows no significant change in the H$\alpha$ emission between 2014 and 2017. 
    
    SN 1961V was originally classified as a ``Type V" supernovae \citep{zwicky_ngc_1964}, and its true nature remains disputed. The progenitor of SN 1961V was one of the brightest stars in NGC 1058. Pre-transient, it had M$_B\approx -$12 mag until brightening to M$_B\approx -$14 mag around a year prior to the peak. In the following months it brightened to M$_B\approx -$16 mag before reaching a peak at M$_B\approx -$18 mag in December 1961. The star's brightness decreased over a few months before briefly plateauing at M$_B\approx -$13 mag and then fading to M$_B\approx -$11.5 mag where it remained for 4 years until dropping below the point of visibility in 1968 (see the various summaries in \citeauthor{goodrich_sn_1989} \citeyear{goodrich_sn_1989}; \citeauthor{humphreys_luminous_1994} \citeyear{humphreys_luminous_1994}; \citeauthor{humphreys__1999}\citeyear{humphreys__1999}; \citeauthor{kochanek_supernova_2011} \citeyear{kochanek_supernova_2011}; \citeauthor{smith_luminous_2011} \citeyear{smith_luminous_2011}).
    
    The light curve of SN 1961V is peculiar and not typical of either an LBV outburst or a supernova. \citet{goodrich_sn_1989} argued that it was best explained by a non-terminal eruption: first an S Doradus phase, where the luminosity is roughly constant but a drop in the stellar temperature makes the source optically brighter in the years prior to the peak, followed by a major LBV outburst in 1961. Many have tried to identify a survivor at the location of SN 1961V calculated by \cite{klemola_precise_1986}. In Figure \ref{fig:field}, star 6 \citep{filippenko_was_1995}, star 11 \citep{van_dyk_possible_2002}, and star 7 \citep{chu_nature_2004} have all been identified as potential survivors or companions to the progenitor based on their red colors and V $>$ 24 mag. Star 7 is presently the preferred candidate due to its H$\alpha$ emission. 
    
    However, there is strong evidence that SN 1961V was an actual, albeit odd, supernova. \citet{branch_radio_1985}, \citet{cowan_radio_1988}, \citet{stockdale_fading_2001}, and \citet{chu_nature_2004} all detected a fading non-thermal radio source, more like a supernova than an eruption, near the location of star 7. A reanalysis of the radio data by \citet{van_dyk_supernova_2005} found a 2$\sigma$ discrepancy between the center of the radio emission and the location of SN 1961V, which they use to argue that the radio emission was not associated with SN 1961V. \citet{smith_luminous_2011} argue that it was an SN based on the transient's energetics. The post-transient spectra showed strong hydrogen emission with velocity widths of about 2000 km s$^{-1}$ \citep{zwicky_ngc_1964,branch_1961_1971} consistent with a Type IIn SN or an eruption. Assuming the ejecta of a non-terminal eruption coalesced into a dusty shell, we would expect to see IR emission from the dust re-radiating the star's light. \citet{kochanek_supernova_2011} analyzed archival \textit{Spitzer} images of SN 1961V and found no evidence of IR emission. However, \citet{van_dyk_its_2012} argue for obscuration by foreground dust which contributes no IR emission. \citet{kochanek_unmasking_2012} pointed out that this solution makes the energetics argument for an SN by \citet{smith_luminous_2011} even more compelling.   
    
    Here we examine both sources with new optical and near-IR HST observations along with continued monitoring of SN 1954J with the Large Binocular Telescope (LBT). Our goal is to determine if these two sources are less optically luminous than their progenitors due to obscuration by dust formed in the transients. We present our data and models in Section \ref{sec:DATA}, a discussion of each event in sections \ref{sec:54J} and \ref{sec:61V}, and a summary in Section \ref{sec:sum}. 
    
\section{DATA and Models} \label{sec:DATA}
Images of SN1954J and SN1961V were taken in October and December of 2013 using WFC3 on HST under program 13477 (PI: Kochanek). Both SN1954J and SN1961V were imaged a total of 12 times in four filters. For SN1954J, the exposures were 3 $\times$ 430s in \textit{F475W}, 3 $\times$ 430s in \textit{F814W}, 2 $\times$ 699s + 2 $\times$ 49s in \textit{F110W}, and 2 $\times$ 799s in \textit{F160W}. For SN1961V, the exposures were 3 $\times$ 381s in \textit{F475W}, 3 $\times$ 381s in \textit{F814W}, 2 $\times$ 499s + 2 $\times$ 99s in \textit{F110W}, and 2 $\times$ 799s in \textit{F160W}. All of the individual exposures were dithered to
    
\begin{figure}
\centering
   \includegraphics[scale=0.6,trim={0 1.7cm 0 0},clip]{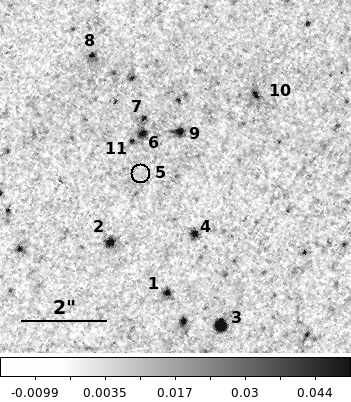} 
   \includegraphics[scale=0.6,trim={0 1.7cm 0 0},clip]{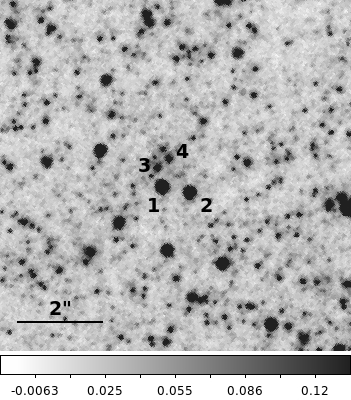}
 \caption{\textit{HST} WFC3 \textit{F814W} images of the regions around SN 1961V (top) and SN 1954J (bottom). Stars 7 and 4 are believed to be associated with SN 1961V and SN 1954J, respectively. The stars are labeled following \citet{filippenko_was_1995} for SN 1961V and \citet{van_dyk_supernova_2005} for SN 1954J. Star 5 from \citet{filippenko_was_1995}, whose position is indicated by the labeled circle, is not visible in any of our data.}
 \label{fig:field}
\end{figure}

\noindent{control for hot pixels and cosmic rays. The pixel scales of the WFC3/UVIS and IR detectors are 0\farcs04/pix and 0\farcs13/pix respectively. Figure \ref{fig:field} shows the \textit{F814W} images of 8'' regions around SN 1954J and SN 1961V. The stars are labeled according to \citet{filippenko_was_1995} for SN 1961V and \citet{van_dyk_supernova_2005} for SN 1954J. Star 5 from \citet{filippenko_was_1995} near SN1961V is not visible in any of our filters.}

  We processed the data using DOLPHOT \citep{dolphin_dolphot:_2016}, separately optimizing the parameters for the UVIS and IR data following the procedures laid out in the manual\footnote{http://americano.dolphinsim.com/dolphot/}. DOLPHOT simultaneously fits point sources to each image using the appropriate PSF for each filter, providing magnitudes and parameters such as sharpness and crowding. All magnitudes are on the Vega scale and presented in Table \ref{tab:phot}. Our magnitudes for SN 1954J are consistent with those in \citet{humphreys_tale_2017}, who analyzed the same data.

	We adopt a distance of 3.3 Mpc to NGC 2403 \citep{freedman_distances_1988} and Galactic reddening of $E(B-V)$ = 0.04 mag \citep{schlegel_maps_1998} and correct for both prior to any modeling. We adopt a distance of 10 Mpc to NGC 1058 \citep{boroson_distribution_1981} and Galactic reddening of $E(B-V)$ = 0.06 mag \citep{schlegel_maps_1998} and again make the requisite reddening corrections before modeling. We assume a minimum error of 10\% on the photometry. We model the stars' spectral energy distributions (SEDs) in two ways: a simple model with additional foreground extinction by an $R_V$ = 3.1 \citet{cardelli_relationship_1989} extinction law and with DUSTY \citep{ivezic_self-similarity_1997,ivezic_user_1999,elitzur_dusty_2001}. In both cases we model the stars using \citet{castelli_new_2004} model atmospheres convolved with the appropriate filter functions. The parameters and their uncertainties are determined using Markov Chain Monte Carlo (MCMC) methods. We do not include the \textit{Spitzer} mid-IR data analyzed in \citeauthor{kochanek_unmasking_2012} (2011, 2012) in our SEDs because no IR emission was observed at the site of SN 1961V and only upper limits were found at the site of SN 1954J.

 The simple model minimizes the fit statistic 
\begin{equation} 
\begin{split}
    \chi^2 = \sum_i \Big(\log{L_i}-\log{L_i^{mod}}(L_*, T_*)\\
    +0.4R_\lambda E(B-V)\Big)^2/\sigma_i^2,
\end{split}
\end{equation}
where $L_i$ and $L_i^{mod}$ are the observed and model band luminosities ($\lambda L_\lambda$), to estimate the total luminosity $L_*$, temperature $T_*$, and additional foreground reddening $E(B-V)$ beyond the Galactic contribution. The $\chi^2$ is calculated and minimized in log space. DUSTY solves for radiation transfer from the star through a dusty shell. We embed DUSTY in an MCMC driver \citep{adams_losss_2015} that allows for a dusty circumstellar shell plus a variable amount of foreground extinction. This model has the V-band optical depth $\tau_V$ and dust temperature $T_d$ as additional parameters. We again optimize the same fit statistic (Equation 1) but in this model, the band luminosities, $L_i^{mod}(L_*, T_*, \tau_V, T_d)$ have more parameters. We fix the dust temperature $T_d$ at 50 K for all our models since the optical and near-IR bands we use have no contribution from the dust emission. For all DUSTY models we treat the dust as purely silicate as this is the type of dust expected from massive stars \citep{speck_dust_2000}. Changes in dust composition have only modest effects on our results (see \citeauthor{kochanek_unmasking_2012} \citeyear{kochanek_unmasking_2012} for further discussion on the effects of dust composition).

Circumstellar absorption due to dust formed in the transient requires that the dust optical depth is time variable (see \citeauthor{kochanek_unmasking_2012} \citeyear{kochanek_unmasking_2012}). The dust forms as the ejecta cools to form a dusty circumstellar shell. As time passes, the optical depth drops as $\tau = \tau_0(t_0/t)^2$ due to its geometric expansion. If the shell fragments or becomes clumpy, the drop in optical depth will accelerate. A clump directly in our line of sight does not prevent a steady drop in the effective optical depth because most of the escaping emission is scattered light from a broad region across the shell. As the optical depth drops, we should see the surviving star brighten with time, as is the case with $\eta$ Carinae (see \citeauthor{humphreys_luminous_1994} \citeyear{humphreys_luminous_1994}). That $\eta$ Carinae had an extended period of constant optical flux means that it must have also been forming a dusty, optically thick wind in that period (\citeauthor{kochanek_unmasking_2012} \citeyear{kochanek_unmasking_2012}). A constant optical depth requires a steady dust forming wind and emission by hot dust, which is not seen for either SN 1954J or SN 1961V (\citeauthor{kochanek_supernova_2011} \citeyear{kochanek_supernova_2011}, \citeyear{kochanek_unmasking_2012}).
   
   Consider a source with intrinsic band luminosity L inside a dusty shell of optical depth $\tau_0$ at time $t_0$. The observed band luminosity is then 
   \begin{equation}
   L_{obs}(t)=Le^{-\tau_0(t_0/t)^2}.
   \end{equation}
If we detect the source at two epochs, $t_0$ and $t_1$, we can determine the optical depth without reference to the object's SED or the type of dust since 
\begin{equation}
\frac{L_{obs}(t_1) - L_{obs}(t_0)}{L_{obs}(t_0)} = e^{\tau_0(1-(t_0/t_1)^2)} - 1.
\end{equation}
Given the optical depth $\tau_0$, we then know the true luminosity. Detection of a change in luminosity determines $\tau_0$, and limits on changes in luminosity set upper bounds on $\tau_0$. 

Crowding, particularly in ground-based observations, means it is only possible to measure changes in luminosity. In particular, difference imaging eliminates crowding, but all you measure is $\Delta L(t_1,t_0) = L_{obs}(t_1) - L_{obs}(t_0)$. Given a long enough light curve, one can still  determine both L and $\tau_0$, but our observations have too short a time baseline to do so. If a change in luminosity is observed, then there is a joint constraint on $L$ and $\tau_0$. An upper limit on the change in luminosity $\Delta L_{max}$, provides an upper limit on the luminosity of the source,
\begin{equation}
L < \frac{\Delta L_{max}}{e^{-\tau_0(t_0/t_1)^2}-e^{-\tau_0}},
\end{equation}
as a function of the optical depth.

For SN 1954J, we have HST observations separated by roughly 10 years, with $t_0 \approx$ 50 and $t_1 \approx$ 60, corresponding to a fractional change in optical depth of $t_0^2/t_1^2 \approx 0.7$. We also have R-band LBT data covering a similar baseline with $t_0 \approx$ 54 and $t_1 \approx$ 64 so again  $t_0^2/t_1^2 \approx 0.7$. For SN 1961V we have HST observations separated by roughly 20 years, with $t_0 \approx 40$ and $t_1 \approx 60$, so the fractional change in the optical depth, $t_0^2/t_1^2 \approx 0.4$, is large. This will make it difficult to invoke a significant optical depth without also requiring an intrinsically low luminosity source.
    
Finally, both star 4 near SN 1954J and star 7 near SN 1961V have strong, broad H$\alpha$ emission, which \citet{van_dyk_its_2012} and \citet{humphreys_tale_2017} use to argue that both transients had survivors. For a fully ionized, constant density, thin hydrogen shell of mass $M$ and radius $R = vt$, the H$\alpha$ recombination luminosity is
\begin{equation}
   L_{H\alpha} = \frac{ M^2 \alpha_{H\alpha}E_{H\alpha}}{4\pi \Delta R^3 m_p^2},
\end{equation} 
where $\alpha_{H\alpha}$ $\simeq$ 10$^{-13}$ cm$^3$ s$^{-1}$, $E_{H\alpha} = $ 1.89 eV, $m_p$ is the proton mass, $\Delta$ is the fractional shell thickness, $v$ is the expansion velocity, and $t$ is the expansion time. This assumes that the outbursts were single events with no continuous outflows. For SN 1961V, this assumption follows a previous analysis by \citet{van_dyk_its_2012}. For SN 1954J, we consider the possibility of dense stellar wind below. Note that the H$\alpha$

 \begin{table*}
  \centering
  \caption{\textit{HST} photometry of stars near SN 1954J and SN 1961V}
  \label{tab:phot}
  \begin{threeparttable}
  \begin{tabular}{llcccccccc}
  \hline
  Event & Object & \textit{F475W} & err & \textit{F814W} & err & \textit{F110W} & err &     \textit{F160W} & err\\
  \hline
  SN 1954J & Star 1 & 24.157 & 0.021 & 20.714 & 0.006 & 19.573 & 0.002 & 18.618 & 0.002\\
  {} & Star 2 & 24.101 & 0.020 & 20.820 & 0.006 & 19.710 & 0.002 & 18.729 & 0.002\\
  {} & Star 3 & 23.190 & 0.012 & 23.073 & 0.021 & 23.109 & 0.018 & 22.940 & 0.034\\
  {} & Star 4\tnote{*} & 24.084 & 0.022 & 22.630 & 0.017 & 22.023 & 0.008 & 21.591 & 0.012\\
  \hline
  SN 1961V & Star 6 & 26.399 & 0.089 & 23.752 & 0.033 & 22.695 & 0.012 & 21.823 & 0.012\\
  {} & Star 7\tnote{*} & 25.808 & 0.058 & 25.198 & 0.103 & 24.418 & 0.045 & 23.831 & 0.056\\
  {} & Star 11 & 27.946 & 0.302 & 25.218 & 0.089 & 24.174 & 0.035 & 23.310 & 0.036\\
  \hline
  \end{tabular}
  \begin{tablenotes}
    \item[*] The most likely eruption survivor
  \end{tablenotes}
  \end{threeparttable}
\end{table*}

\begin{table*}
\centering
\caption{Best fit DUSTY models}
\label{tab:dusty}
\begin{threeparttable}
\begin{tabular}{lllllll}
\hline
Event & Object & $T_{eff}$ & log L$_*$ & E(B$-$V) & $\chi^2$/dof & Priors\\
{} & {} & (K) & {} & (mag)\\
\hline
SN 1954J & Star 1 & $3730^{+3230}_{-190}$ & $4.63^{+0.89}_{-0.10}$ & $0.25^{+1.20}_{-0.19}$ & 0.077 & {}\\
{} & Star 2 & $3810^{+12760}_{-240}$ & $4.59^{+2.03}_{-0.11}$ & $0.26^{+1.60}_{-0.20}$ & 0.049 & {}\\
{} & Star 3 & $12470^{+1480}_{-3280}$ & $4.18^{+0.98}_{-0.39}$ & $0.14^{+0.19}_{-0.11}$ & 0.392 & {}\\
{} & Star 4\tnote{*} & $6590^{+3290}_{-1540}$ &  $4.01^{+0.62}_{-0.34}$ & $0.39^{+0.41}_{-0.35}$ & 0.047 & No $T_*$ prior\\
{} & Star 4\tnote{*} & $5230^{+290}_{-350}$ & $3.71^{+0.06}_{-0.07}$ & $0.09^{+0.07}_{-0.08}$ & 0.468 & $T_*$ prior\\
\hline
SN 1961V & Star 6 & $4740^{+760}_{-880}$ & $4.58^{+0.19}_{-0.31}$ & $0.56^{+0.23}_{-0.52}$ & {0.000} & {}\\
{} & Star 7\tnote{*} & $5860^{+10570}_{-180}$ & $3.78^{+1.40}_{-0.00}$ & $0.04^{+0.70}_{-0.01}$ &  8.228 & All bands\\
{} & Star 7\tnote{*} & $7510^{+830}_{-770}$ & $3.78^{+0.06}_{-0.06}$ & $0.06^{+0.03}_{-0.05}$ &  0.00 & {Optical bands only}\\
{} & Star 11 & $4830^{+1010}_{-950}$ & $4.01^{+0.25}_{-0.30}$ & $0.59^{+0.32}_{-0.50}$ & 0.046 & {}\\
\hline
\end{tabular}
\begin{tablenotes}
\item{*} The most likely eruption survivor.
\item NOTE - Errors reflect the 90\% confidence interval. The extinction represents additional absorption beyond the Galactic contribution.
\end{tablenotes}
\end{threeparttable}
\end{table*}  

\begin{table*}
\centering
\caption{2-star models of star 4 near SN 1954J }
\label{tab:mods}
\begin{threeparttable}
\begin{tabular}{lllllll}
\hline
model type & T$_H$ & T$_C$ & log L$_H$ & log L$_C$ & E(B$-$V) & $\chi^2$/dof\\
{} & (K) & (K) & {} & {} & (mag) & {}\\
\hline
2 stars fixed T$_H$ and T$_C$ & 10000 & $5180^{+470}_{-450}$ & $0.30^{+2.76}_{-0.16}$ & $3.70^{+0.11}_{-0.10}$ & $0.08^{+0.15}_{-0.08}$ & 0.233\\
2 stars fixed T$_H$ and T$_C$ & 15000 & $5190^{+480}_{-460}$ & $0.40^{+3.22}_{-0.21}$ & $3.70^{+0.12}_{-0.09}$ & $0.08^{+0.18}_{-0.08}$ & 0.233\\
2 stars fixed T$_H$ and T$_C$ & 20000 & $5200^{+440}_{-460}$ & $1.00^{+2.57}_{-0.82}$ & $3.70^{+0.11}_{-0.10}$ & $0.08^{+0.13}_{-0.08}$ & 0.233\\
2 stars fixed T$_H$ and T$_C$ & 25000 & $5180^{+470}_{-430}$ & $1.06^{+2.76}_{-0.90}$ & $3.70^{+0.11}_{-0.09}$ & $0.08^{+0.15}_{-0.07}$ & 0.233\\
\hline
\end{tabular}
\begin{tablenotes}
\item NOTE - Here H refers to the hot star and C to the cooler star. Errors reflect the 90\% confidence interval.
\end{tablenotes}
\end{threeparttable}
\end{table*}

\noindent{luminosity is not constant but decays as $L_{H\alpha} \propto R^{-3} \propto t^{-3}$ with time. If the H$\alpha$ emission comes from photoionizing the ejecta, then the luminosity provides an estimate of the mass, with
\begin{equation}
M \simeq 0.1 v_{3}^{3/2} t_{50}^{3/2} L_{36}^{1/2} \Delta_{0.1}^{1/2} M_\odot,
\end{equation}
where $t = 50t_{50}$ years, v = 1000v$_3$ km s$^{-1}$, $L_{H\alpha}=10^{36}L_{36}$ erg s$^{-1}$, and $\Delta = 0.1\Delta_{0.1}$.}

Any dust formed in the ejecta is mixed with the hydrogen, so the dust optical depth of the ejecta is 
\begin{equation}
\tau = \frac{M\kappa}{4\pi R^2}
\end{equation}
where $\kappa$ is the dust opacity. We can solve for the total H$\alpha$ luminosity in terms of the optical depth $\tau$ and ejection velocity $v$ to find that 
 \begin{equation}
 L_{H\alpha}=\frac{4 \pi v t \tau^2 \alpha_{H\alpha} E_{H\alpha}}{\Delta \kappa^2 m_p^2}.
 \end{equation}
 Note that this scaling only depends on dust properties like size distribution and composition only through the opacity. This does not account for dust absorption in the shell. While there is no analytic expression for a shell, the observed luminosity from a cube of side 2$R$ is smaller by (1$-e^{-2\tau})/2\tau$, which goes to 1/{2$\tau$} in the limit of large optical depth. The $\tau^{-1}$ scaling is generic because you only see radiation from the regions where the optical depth to the observer is $\tau \lesssim 1$ which is a fraction $\tau^{-1}$ of the overall volume. Thus, in the high optical depth limit
 \begin{equation}
 L_{H\alpha} \simeq \frac{4 \pi v t \tau \alpha_{H\alpha} E_{H\alpha}}{\Delta \kappa^2 m_p^2}.
 \end{equation}
Solving for $\tau$ we find,
\begin{equation}
\tau = 0.07 \kappa_2 t_{50}^{-1/2}{v_3}^{-1/2}L_{36}^{1/2}\Delta_{0.1}^{1/2}
\end{equation}
for $\tau \leq 1$, and
\begin{equation}
\tau \simeq 0.00 \kappa_2^2 t_{50}^{-1}{v_3}^{-1}L_{36}\Delta_{0.1}
\end{equation}
for $\tau \geq 1$, where $\kappa=100\kappa_2$ cm$^2$ g$^{-1}$. If the material is in a uniform sphere, then the optical depth is larger by $(3/\Delta)^{1/2}$ or $(3/\Delta)$ for the low and high $\tau$ cases, respectively. We also note that the dust destruction time in such a photoionized nebula is $\gg 10^2$ years (see, e.g., \citeauthor{draine_grain_1995} \citeyear{draine_grain_1995}). Dust formed in the ejecta will continue to be present if the ejecta is photoionized at a later time.

Finally, we consider a scenario put forth by \citet{humphreys_tale_2017}, where the H$\alpha$ emission comes from a dense stellar wind of a binary companion. The H$\alpha$ luminosity of the wind is 
\begin{equation}
  L_{H\alpha} = {\dot{M}^2 \alpha E_{H\alpha} \over 4 \pi v_w^2 \mu^2 m_p^2 R_* }.
\end{equation}
where $\dot{M}$ is the mass loss rate, $v_w$ is the wind speed, and $R_*$ is the stellar radius. The Thomson optical depth of the wind is
\begin{equation}
    \tau_T = { \dot{M} \kappa_T \over 4 \pi R_* v }
\end{equation} 
where the Thomson opacity is $\kappa_T \simeq 0.5$~cm$^2$/g. \citet{humphreys_tale_2017} want $\tau_T \simeq 1$ to explain the shape of the line profile of SN 1954J. The H$\alpha$ luminosity is then 
\begin{equation}
  L_{H\alpha} = {4 \pi R_* \alpha E_{H\alpha} \tau_T^2 \over \mu^2 m_p^2 \kappa_T^2 },
\end{equation}
in terms of its Thomson optical depth. Finally, we can express the stellar radius $R_*^2 = L_*/4 \pi \sigma T_*^4$ in terms of the stellar luminosity $L_*$ and effective
temperature $T_*$, to find that
\begin{equation}
  L_{H\alpha} = 84 L_{*2}^{1/2} T_{*2}^{-2} \tau_T^2 L_\odot
\end{equation}
for $L_*=100 L_{*2} L_\odot$, $T_* = 20000 T_{*2}$~K, $\kappa=0.5$~cm$^2$/g, $\alpha = 10^{-13}$~cm$^3$/s
and $\mu = 1$. 

Adding dusts obscuration has an interesting consequence due to the $L_{H\alpha} \propto L_{*2}^{1/2}$ scaling of Equation 15.  Let $L_{H\alpha 0}$ and $L_{*0}$ be the H$\alpha$ luminosity
inferred from the line flux and the stellar luminosity inferred from Equation 15 assuming there is no dust. Suppose we add a dust optical depth $\tau_\alpha$ at the wavelength of H$\alpha$. Then the true line luminosity is $L_{H\alpha} = L_{H\alpha0}\exp(\tau_\alpha)$ and we must have $L_* = L_{*0}\exp( 2 \tau_\alpha)$. This means that the expected flux of the star, $F_* \propto L_* \exp(-\tau_\alpha) \propto L_{*0}\exp(\tau_\alpha)$ actually increases rather than decreases as you try to obscure the star behind dust because of how the H$\alpha$ luminosity depends on $L_*$ when you hold the Thomson optical depth of the wind constant. Note that if we consider the flux of the star near H$\alpha$ this is completely independent of dust properties.

\section{SN 1954J in NGC 2403} \label{sec:54J}
\begin{figure*}
  \hfill\includegraphics[scale=0.48,trim={0.75cm 0cm 0cm 1.5cm},clip]{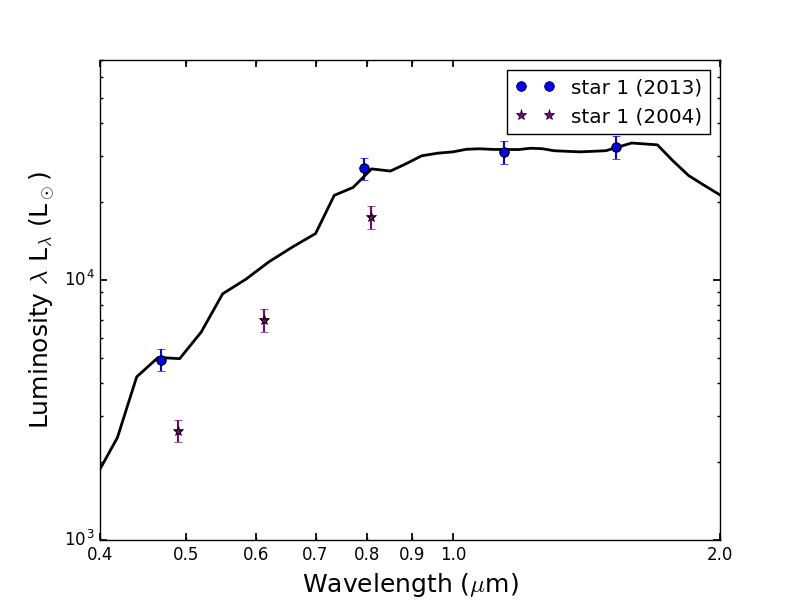}\hspace*{\fill}
  \hfill\includegraphics[scale=0.48,trim={0.75cm 0cm 0cm 1.5cm},clip]{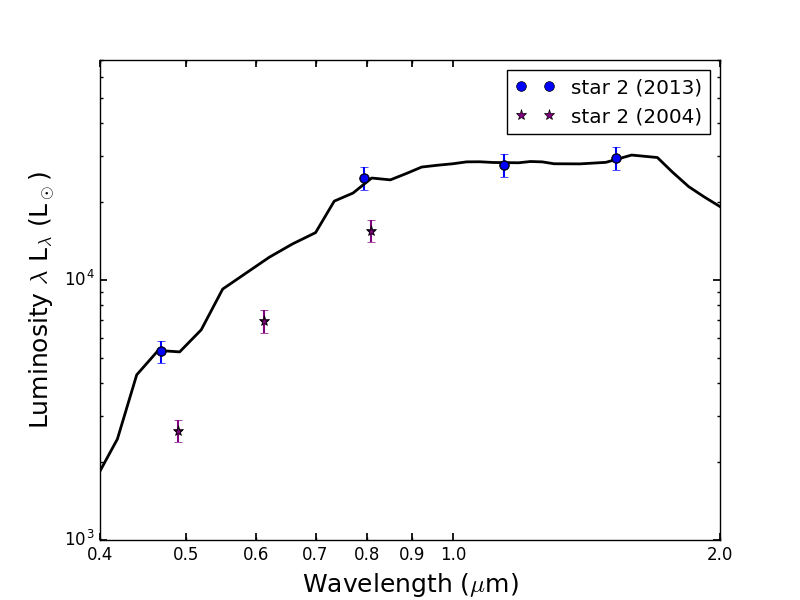}\hspace*{\fill}
 
  \hfill\includegraphics[scale=0.48,trim={0.75cm 0cm 0cm 1.5cm},clip]{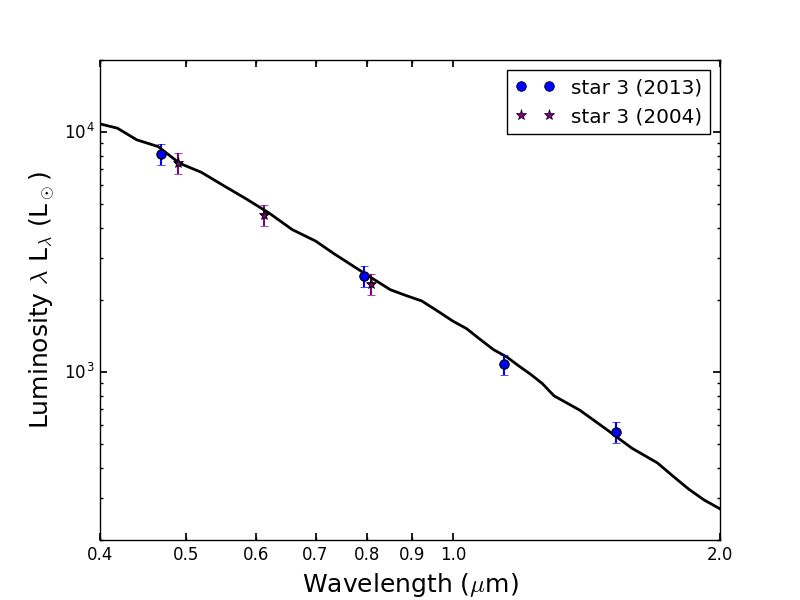}\hspace*{\fill}
  \hfill\includegraphics[scale=0.48,trim={0.75cm 0cm 0cm 1.5cm},clip]{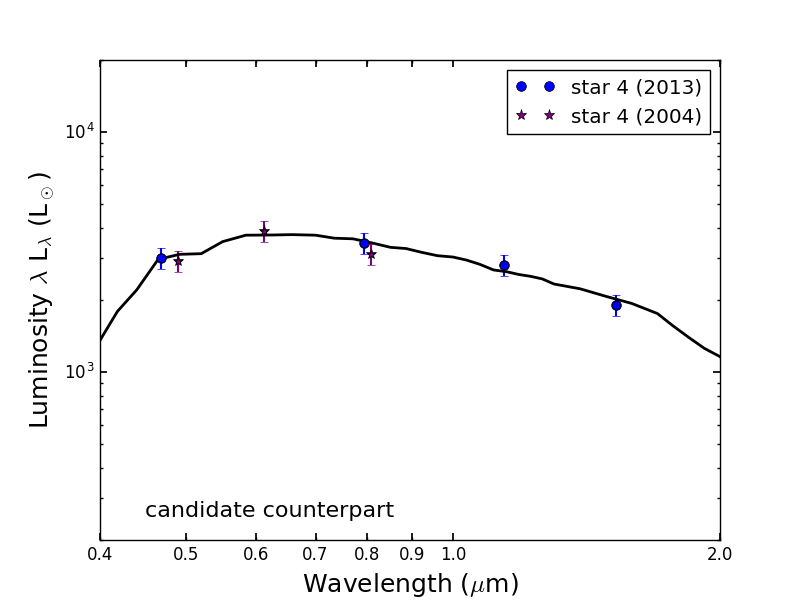}\hspace*{\fill}
  
  \caption{The SEDs of stars 1, 2, 3, and 4 near SN 1954J for our current data (circles) and the archival ACS/WFC \textit{F475W}, \textit{F606W}, and \textit{F814W} photometry (stars) from \citet{humphreys_tale_2017}. The model SED for star 4 includes the temperature prior $T_* = 5000 \pm 500$ K. The \textit{F475W} and \textit{F814W} points are offset by $\pm$ 0.01 $\mu$m from the mean filter wavelength so that they do not overlap. These models assume the foreground extinction for the best fit single star models (i.e., Table \ref{tab:dusty})}
 \label{fig:SEDJ}
\end{figure*}

All four stars in the environment of SN 1954J are well fit as single stars. Table \ref{tab:dusty} lists the best fit parameters for each star and Figure \ref{fig:SEDJ} shows each star's SED after being corrected for the model foreground extinction. We assumed the same foreground extinction for the archival and current band luminosities. \citet{humphreys_tale_2017} argue that an additional A$_V$ of 0.8-0.9 beyond Galactic is necessary to properly fit each star, which is consistent with our findings. Stars 1 and 2 appear to be red giants and star 3 is an A or B star. 

The candidate counterpart to SN 1954J, star 4, has an intermediate temperature. \citet{humphreys_tale_2017} note that their blackbody fit gives T $\approx$ 6600 K while their spectroscopic estimate is 5000 K. Our fits based on model atmospheres have uncertainties large enough to be consistent with either estimate (T$_* \approx$ 6600$^{+3300}_{-1500}$ K). If we constrain its temperature to lie within 10\% of their spectroscopic temperature by including a penalty of $(T_*-5000)^2$/$\sigma_T^2$ with $\sigma_T$ = 500 K in the fit statistic, the difference in the fit statistic between the two models $(\Delta \chi^2 = 0.42)$ shows that the SED is consistent with the spectroscopic temperature (see Table \ref{tab:dusty}).

We again modeled star 4 with DUSTY, this time including the temperature prior and allowing the circumstellar optical depth ($\tau_v$) to vary. If we allow both the foreground and circumstellar extinction to vary we, not surprisingly, find degenerate solutions. If we fix the the foreground extinction to $E(B-V) = 0.25$ to match the neighboring stars, we find $\tau_V = 0.03^{+0.73}_{-0.03}$, consistent with the results from \citet{kochanek_unmasking_2012}, while if we fix $E(B-V)=0$, we find $\tau_V = 1.26^{+0.75}_{-1.08}$.
    
\citet{humphreys_tale_2017} ultimately propose that the system consists of two stars, a ``cold" star with $T_C\simeq$ 5000 K and $\log L_* \simeq$ 4.6 and a ``hot" star with $T_H \simeq$ 20000 K and $\log L_* \simeq$ 5.3. To explore this model further, we constrained the temperature of the cooler star by our temperature prior of $T_C = 5000 \pm 500$ K and consider fixed temperatures for the hot star of $T_H = 10000, 15000, 20000,$ and $25000$ K. The luminosities of the two stars were free to vary, but with a conservative penalty in the fit statistic of $(L_H/L_C)^2$ for the \textit{F475W} and \textit{F814W} band luminosities if $L_H>L_C$. We know from the \citet{humphreys_tale_2017} spectrum that the observed optical flux has to be dominated by the cooler star, so we (conservatively) should not allow the hot star to dominate the optical emission. Both stars are subject to the same extinction.
   
The results for these models are shown in Table \ref{tab:mods}. One can never rule out the presence of a hot star, but all of our models require the hot star to have a negligible luminosity which cannot produce sufficient ionizing photons to explain the line emission. If we try to put in a hot star with a luminosity of $L_H = 10^{5.3} L_\odot$ and $T_H$ = 20000 K, as proposed by \citet{humphreys_tale_2017}, we find unacceptable fits with $\chi^2 = 4.4$.

While we agree with \citet{humphreys_tale_2017} on the amount of dust, either foreground or circumstellar, needed to reproduce the stellar colors, \citet{humphreys_tale_2017} then add $A_V$ = 2.5 of gray dust to allow star 4 to be intrinsically more luminous. They argue that the grains formed in the eruption might be large enough to be effectively gray. Such dust cannot be identified in the SED fits, but it is constrained by the time variability of the source and the H$\alpha$ luminosity.

As discussed in Section \ref{sec:DATA}, we can use variability to constrain the optical depth of any dust formed in the transient. For SN 1954J, we can do this in two ways. First, we have archival R band images from the Large Binocular Telescope

\noindent{from March 2008 through May 2018 \citep{gerke_search_2015,adams_search_2017}. We see no variability at its location. To set a limit on the variability, we selected a nine-point grid with 2" offsets from the location of SN 1954J and measured the brightness in each epoch at each point over the 9 year time span. The average slope at the site of SN 1954J was 30 $\pm$ 90 counts/year. Taking our 1$\sigma$ upper limit as the estimate, the average slope was 120 counts/year, which at 0.36 L$_\odot$/count, corresponds to approximately 43 L$_\odot$/year, or an upper limit of $\Delta L = $ 430 L$_\odot$ over the 10 year baseline.}
  
We know from the HST data and the SED fits to star 4 that the observed R band luminosity in 2013.9 was $L_R = 7527 L_\odot$. The LBT data span from 2008.5 ($t_a \simeq 54$ years) to 2018.4 ($t_b \simeq 64$ years) so the change in luminosity should be
\begin{equation}
  \Delta L = L_R e^{\tau_{HST}}[e^{-\tau_{HST}(t_{HST}/t_b)^2}-e^{-\tau_{HST}(t_{HST}/t_a)^2}],
\end{equation}
where $t_{HST} \simeq 60$ years. The observed lack of variability implies an upper limit on the R band optical depth of $\tau_{HST} < 0.16$ at R band at the time of the HST observations. The second approach is to compare our new HST photometry to that from \citet{humphreys_tale_2017}, who present photometry updated from the initial results reported in \citet{van_dyk_supernova_2005}. They report \textit{F475W} and \textit{F814W} magnitudes of 24.11 mag and 22.73 mag, respectively, as shown in Figure \ref{fig:SEDJ}. These correspond to changes of 0.03 mag and 0.10 mag, both in the sense of the star brightening. If we treat these as upper limits on the star becoming brighter, Equation 3 implies $\tau_{F475W} < 0.07$ and $\tau_{F814W} < 0.25$ at the time of the first HST epoch near the end of 2013. The lack of variability strongly implies that star 4 cannot be surrounded by a dusty expanding medium with any significant optical depth.

Finally, \citet{van_dyk_supernova_2005} argued that star 4 was the most likely survivor of SN 1954J due to its strong H$\alpha$ emission. Like the lack of optical variability, the lack of H$\alpha$ variability is difficult to reconcile with material ejected in 1954. Recall that H$\alpha$ emission coming from the ionized ejecta should fall off like $t^{-3}$ (Equation 5). \citet{humphreys_tale_2017} report no significant change in the H$\alpha$ flux between 2014.0  and 2017.1, a period over which we would expect the H$\alpha$

  \begin{figure*}
  \centering
 \hfill\includegraphics[scale=0.48,trim={0.75cm 0cm 0cm 1.5cm},clip]{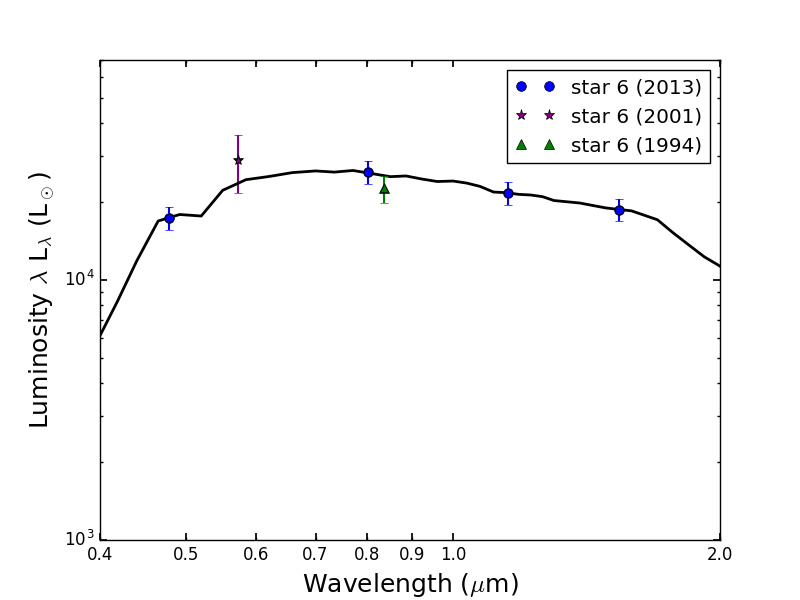}\hspace*{\fill}
  \hfill\includegraphics[scale=0.48,trim={0.75cm 0cm 0cm 1.5cm},clip]{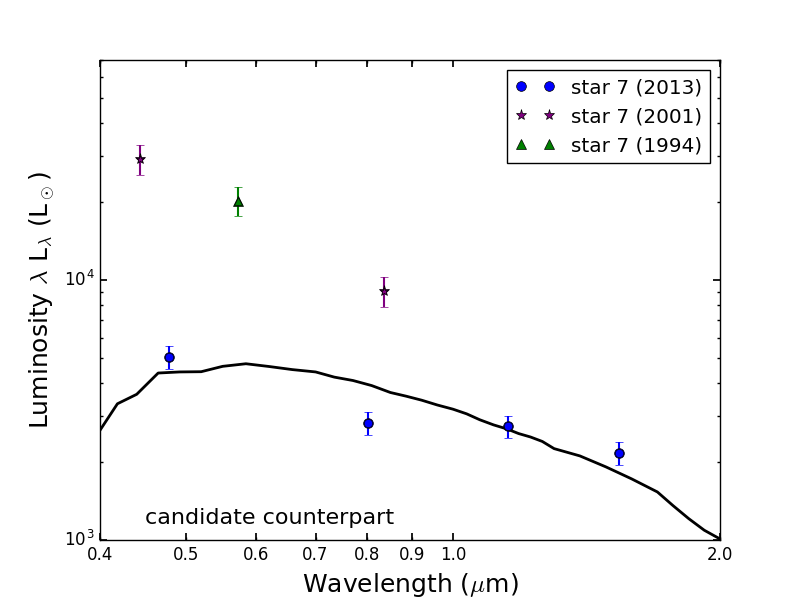}\hspace*{\fill}
  
 \hfill\includegraphics[scale=0.48,trim={0.75cm 0cm 0cm 1.5cm},clip]{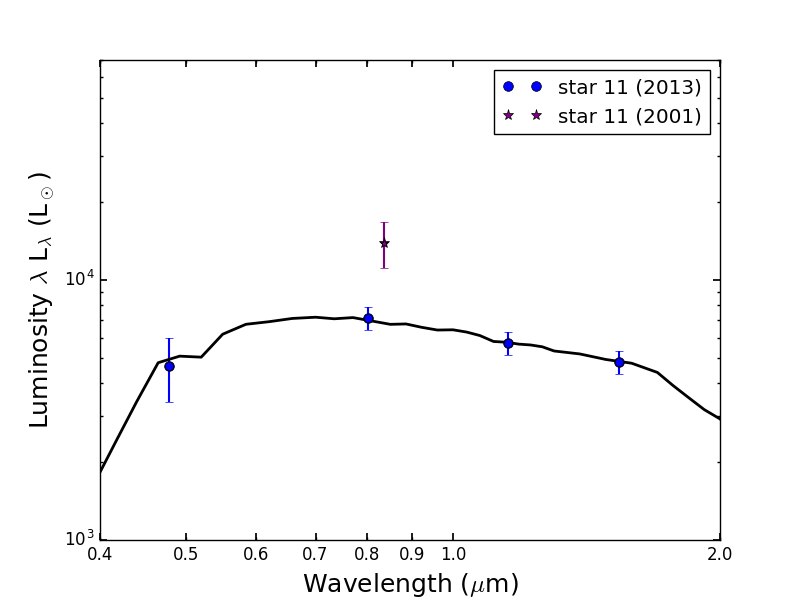}\hspace*{\fill}
  \hfill\includegraphics[scale=0.48,trim={0.75cm 0cm 0cm 1.5cm},clip]{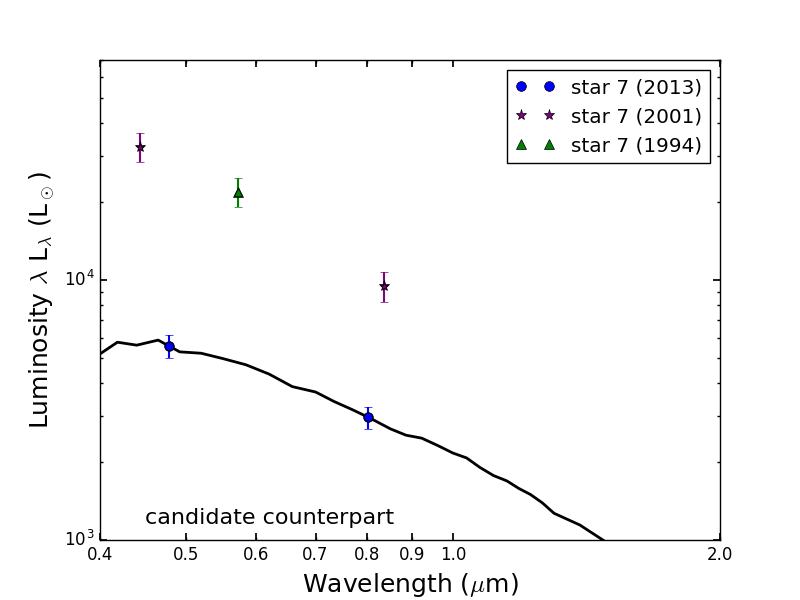}\hspace*{\fill}
 
 \caption{The measured SED of stars 6 (top left), star 11 (bottom left), and candidate counterpart star 7 (right) near SN 1961V with our current data (circles) and the archival photometry (stars and triangles) from \citet{van_dyk_possible_2002}. In the top right panel we show the best fit to the full SED while the bottom panel shows the fit to just the optical bands. Star 7 appears to have significantly faded but this may be an artifact (see text).}
 \label{fig:SED}
  \end{figure*}
  
\noindent{flux to drop by $\sim 14\%$. More significantly, the H$\alpha$ flux found by \citet{humphreys_tale_2017} appears to be significantly greater than that reported by \citet{van_dyk_supernova_2005} at 2002.9, a period over which the H$\alpha$ luminosity should have dropped by a factor of $\sim 2$. \citet{van_dyk_supernova_2005} did report poor seeing conditions, which may account for the difference in luminosity.}
  
We can also use the observed $H\alpha$ luminosity of $L_{H\alpha}$ = $1.3 \times 10^{36}$ erg s$^{-1}$ \citep{humphreys_tale_2017} to constrain the dust optical depth. Following the arguments in Section \ref{sec:DATA}, we find that the amount of dust associated with the H$\alpha$ emission can be at most
\begin{equation}
\tau \approx 0.073\kappa_2t_{50}^{-1/2}v_3^{-1/2}\Delta_{0.1}^{1/2},
\end{equation}
in the low optical depth limit and
  
\begin{equation}
\tau \approx 0.004 \kappa_2^2 t_{50}^{-1} v_3^{-1}\Delta_{0.1},
\end{equation}
in the high optical depth limit. Here we set $t_{50}=1.26$, corresponding to the optical depth in 2017. Both of these results are consistent with our upper limits on circumstellar absorption from the variability, and neither imply a high optical depth. If additional absorption is placed outside the H$\alpha$ emitting region, the limits on $\tau$ increase as $L_{H\alpha}^{1/2}$ ($L_{H\alpha}$) in the low (high) optical depth limit. However, such dust would have to be pre-existing and not formed in the transient.

Finally, in the hot binary companion with a wind scenario used by \citet{humphreys_tale_2017} to explain the broad asymmetric wings of star 4's H$\alpha$ profile with Thompson scattering, Equation 15 means that L$_{H\alpha}$ = 340 L$_\odot$ corresponds to a stellar luminosity of L$_*$ = 1600 L$_\odot$ for a Thompson optical depth of $\tau_T=1$ and no dust extinction.  This already appears to rule out this hypothesis, as it is very difficult to see how a star could emit a fifth of its total luminosity in H$\alpha$ emission. The luminosity is also above our limits for the luminosity of a hot companion (Table \ref{tab:mods}). If we now add extinction, not only are the luminosity limits on a companion violated even further, but the constraint of a constant H$\alpha$ luminosity would actually force the optical continuum flux from the companion upwards rather than downwards, in further conflict with the data. In summary, our SED fits, the lack of variability, and the H$\alpha$ luminosity and its evolution in either a shell or a wind scenario all suggest that star 4 has negligible circumstellar extinction associated with the transient or any present day wind. Aside from the SED fits, these constraints apply even to gray dust.

\section{SN 1961V in NGC 1058} \label{sec:61V}
Table \ref{tab:phot} also includes the photometry of stars 6, 7, and 11 (see Figure \ref{fig:field}) near SN 1961V, where star 7 is believed to be the counterpart of SN 1961V. The SEDs of the stars are shown in Figure \ref{fig:SED}.  Both the archival and current band luminosities have been corrected for foreground Galactic extinction. We modeled the stars with DUSTY, with the results in Table \ref{tab:dusty} and Figure \ref{fig:SED}. Stars 6 and 11 are well fit, but star 7 is not, as might be expected given the SED. The issue appears to be crowding in the WFC3/IR images, where stars 7 and 6 are not well resolved. This is reflected in the DOLPHOT crowding parameter for star 7. Crowding measures how much brighter a star would be in magnitudes for a given filter if all of the stars in the image were not fit simultaneously. Star 7 has crowding corrections of 0.501 mag and 0.712 mag in \textit{F110W} and \textit{F160W} respectively, compared to 0.050 mag and 0.099 mag in \textit{F475W} and \textit{F814W}. This likely makes the near-IR photometry of star 7 unreliable.

Since we were unable to ``fix" the photometry, we first worked to obtain upper limits on the luminosity of star 7. We  calculate the total stellar luminosity normalized to the star's luminosity at \textit{F475W} over a grid of temperature and $E(B-V)$. We let temperature run from 3500 K to 25000 K in increments of 500 K. We let the reddening run from $E(B-V) =$ 0 to 1 in increments of 0.05 mag and from 1 to 2 in increments of 0.1 mag. Given the total luminosity we calculate the band luminosities for the other three filters and set $\chi^2 = \sum_i (L_i/L_{mod})^2$, keeping only the points with $\chi^2 \leq 12$, corresponding to the model exceeding (on average) the flux in each of the other three filters by a factor of two. We then repeat the process, this time normalizing to the luminosity in \textit{F160W}.

\begin{figure}
 \includegraphics[scale=0.48, trim={1cm 0 0 0},clip]{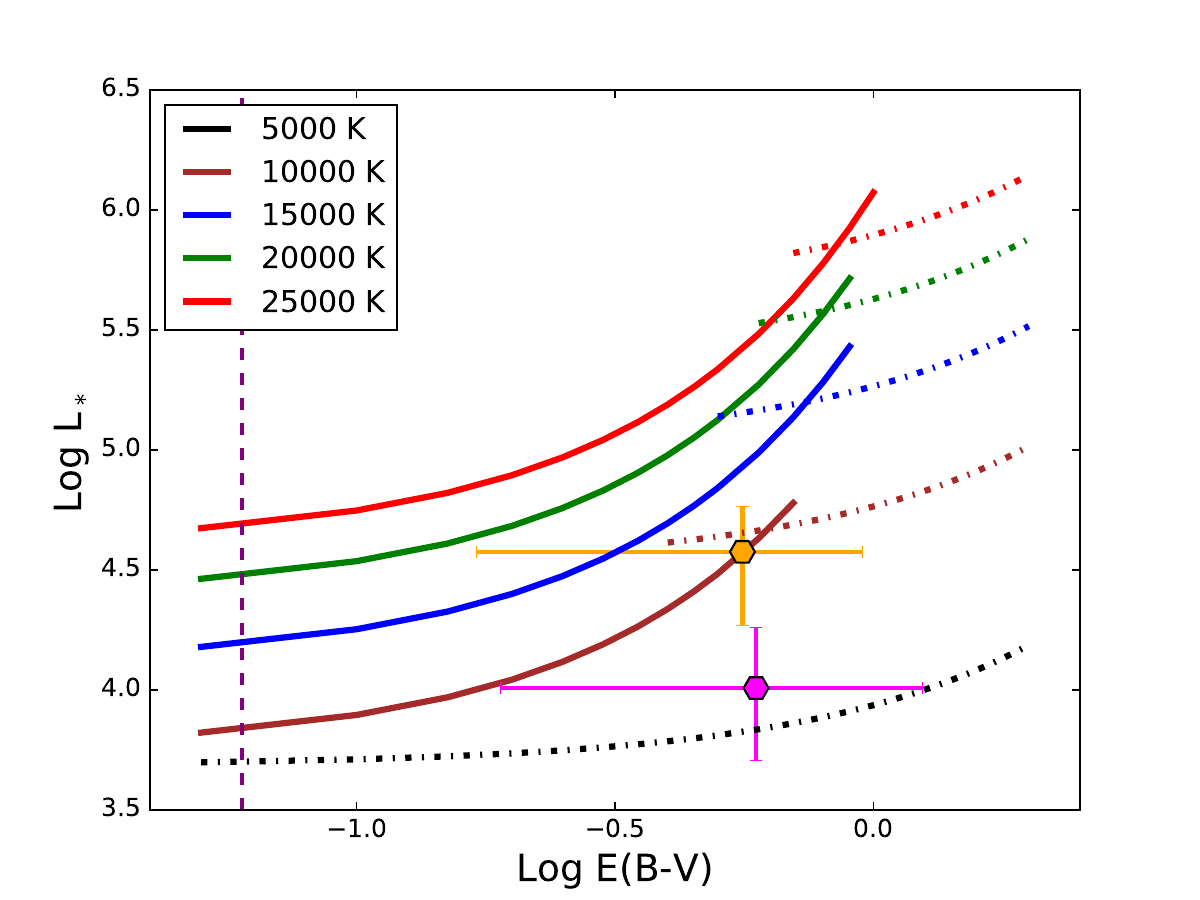}
 \caption{Constraints on the luminosity and E(B$-$V) of star 7 from our photometry normalized at either \textit{F475W} (solid) or \textit{F110W} (dot-dashed). Stars 11 (magenta) and 6 (orange) are shown for comparison. $E(B-V)$ is the extinction in excess of the Galactic contribution.}
 \label{fig:constraint}
\end{figure}

Figure \ref{fig:constraint} shows the allowed luminosities as a function of $E(B-V)$ and fixed temperature for both the blue- and red-normalized models. The progenitor had L$_*$ $\approx$ 10$^{6.4}$ L$_\odot$ \citep{goodrich_sn_1989}, so we expect the survivor to be comparably luminous. Only the hottest stars with the highest reddening come close to reaching the progenitor luminosity. The measured stellar luminosity and $E(B-V)$ for stars 6 and 11 are shown for comparison. In these models, hot stars can be luminous because they need only pass through the \textit{F475W} band point while lying below all the other bands. In the red-normalized case, with the right amount of extinction, the observed SED can be matched to the extincted Rayleigh-Jeans tail of a hot star, allowing the very high total luminosities for the red-normalized case.
  
However, we trust both the \textit{F475W} and \textit{F814W} band luminosities and so should have our models fit both. Unlike the IR data, there are no flags indicating any problems, and star 7 is nicely visible and isolated in both images. If we ignore star 7's near-IR photometry and just model the optical photometry with DUSTY (see Figure \ref{fig:SED}), we find much tighter constraints on the star's temperature, luminosity, and foreground extinction. Table \ref{tab:dusty} shows the results. 

\noindent{The luminosity is required to be nearly three orders of magnitude less than that of the progenitor of SN 1961V.}

Figure \ref{fig:SED} also shows the \textit{F606W} data from 1994 (PI: Illingworth under program 5446) and the \textit{F450W} and \textit{F814W} data from 2001 (PI: Smartt under program 9042) reported in \citet{van_dyk_possible_2002}. Star 7 appears to have systematically faded by 1.7 mag in \textit{F475W}. It would require an increasing optical depth of $\Delta \tau_{F475W}$ = 1.74 to explain this change with dust. In \textit{F814W}, there is a systematic fading of 1.4 mag, which corresponds to a change in optical depth of $\Delta \tau_{F814W}$ = 1.17. For comparison, stars 4, 6, and 9, show no significant changes, and any changes for stars 2 and 11 (Figure \ref{fig:SED}) are far less significant. However, the quality of the old data is poor and the region around SN 1961V is only barely on the detector (see Figures 2 and 3 in \citeauthor{van_dyk_possible_2002} \citeyear{van_dyk_possible_2002}). Examining the images, we had difficulty convincing ourselves that the fading is actually real, but it is at least clear that there is no significant brightening of the source. If we pretend that the change in magnitude is reversed and that the source could have at most brightened by 1.7 and 1.4 mag in \textit{F475W} and \textit{F814W}, we find that $\tau_{F475W} < 1.02$ and $\tau_{F814W} < 0.89$ based on Equation 4. 
    
\citet{van_dyk_its_2012} argue for the existence of a survivor of SN 1961V based partially on the H$\alpha$ emission. If the H$\alpha$ emission comes from the ionized ejecta, then $L_{H\alpha}$ should fall off with time, in this case, by a factor of 4.4 between the last two spectra taken of star 7. We examined the reported H$\alpha$ fluxes from \citet{goodrich_sn_1989} and \citet{chu_nature_2004} looking for signs of variability, but we could not identify any directly comparable estimates of the H$\alpha$ flux.
    
Using the most recent measurement of $L_{H\alpha}$, we can still place limits on the dust optical depth. \citet{van_dyk_its_2012} report that $L_{H\alpha}$ for SN 1961V is $6.5\times10^{36}$ erg s$^{-1}$, implying that
\begin{equation}
  \tau \approx 0.19\kappa_2{v_3}^{-1/2}\Delta_{0.1}^{1/2},
\end{equation}
in the low optical depth limit, and
    \begin{equation}
  \tau \approx 0.04\kappa_2^2t_{50}^{-1}{v_3}^{-1}\Delta_{0.1},
  \end{equation}
in the high optical depth limit using Equations 9 and 10. We have set $t_{50} = 0.82$ corresponding to the optical depth in 2002. As before, adding additional extinction external to the H$\alpha$ emission can weaken these limits as $L_{H\alpha}^{1/2}$ ($L_{H\alpha}$) for the low (high) optical depths. However, as emphasized by \citet{kochanek_unmasking_2012}, adding any additional distant extinction essentially forces the event to be a supernova on energetic grounds. 

We also considered the wind scenario from \citet{humphreys_tale_2017} for SN~1961V.  With L$_{H\alpha}$ = 1700 L$_\odot$, Equation 15 implies L$_*$ = 41000 L$_\odot$ for $\tau_T = 1$, T$_* = 20000$~K and no dust. Reducing T$_*$ to 7500~K to reflect the best fit SED model (optical bands only) gives  L$_*$ = 810 L$_\odot$, although such a star is also too cool to ionize the wind.  As discussed in \ref{sec:DATA} and for SN 1954J, while adding dust increases the stellar luminosity, doing so at fixed H$\alpha$ flux also means that the observed continuum flux of the star increases rather than decreases as the dust optical depth is increased, violating the observed flux limits on the star.
 
\section{Summary} \label{sec:sum}
	SN 1954J and SN 1961V were peculiar events, and the wealth of literature on the two over the past 60 years is a testament to the difficulty of classifying them. The fundamental challenge is that the candidate counterparts have present day optical magnitudes significantly fainter than the progenitors. We know from \textit{Spitzer} observations that neither source can be obscured by hot dust forming in a present day wind (\citeauthor{kochanek_supernova_2011} \citeyear{kochanek_supernova_2011}, \citeyear{kochanek_unmasking_2012}) This leaves three options for reconciling the fluxes if these are the surviving stars. First, the stars can become obscured by cold dust formed in the transient. Second, the star can be comparably luminous but have a much higher temperature so that bolometric corrections allow the optical fluxes to be larger. Third, the star could have become intrinsically fainter. We are currently unaware of any mechanism which would cause the surviving star to become intrinsically fainter, so the latter explanation is the least likely of the three.   
    
    Here we have used four different probes of the optical depth of any dust formed in the transient: SED fits, photometric variability, and the H$\alpha$ luminosity and its evolution. All four probes imply little dust optical depth associated with the transient for both sources, far too little to significantly modify the luminosity of either star. This includes any significant gray opacity as invoked by \citet{humphreys_tale_2017} in order to bring the luminosity of star 4 near SN 1954J back to its pre-transient level. While our SED fits have no sensitivity to gray dust, the time variability and the H$\alpha$ luminosity limits hold regardless of the extinction curve.
    
    The last spectra taken of stars 4 and 7 were in 2017 and 2002 respectively. New spectra of these targets could verify any H$\alpha$ variability. By 2020, we would expect to see any H$\alpha$ luminosity from the ionized ejecta fade by $\sim 13 \%$ for SN 1954J and $\sim 66 \%$ for SN 1961V. While we argue that the data already rule out a significant dust optical depth associated with the ejecta, there is also a problem with the required ejecta mass. For a thin shell, expanding at velocity $v$, the shell mass needed to produce an optical depth of $\tau$ is  
\begin{equation}
M = \frac{4 \pi v^2 t^2}{\kappa} \tau = 1.6 \kappa_2^{-1} v_3^{2} t_{50}^2 \tau M_\odot,
\end{equation}
which means that both SN 1954J and SN 1961V require ejecta masses of $\sim$10$M_\odot$ to have a significant amount of absorption ($\tau \simeq 3$). This is not impossible, as it roughly matches estimates for $\eta$ Carinae (e.g., \citeauthor{morris_discovery_1999} \citeyear{morris_discovery_1999}). Note, however, that the mass implied by the H$\alpha$ luminosity (Equation 6) is just 0.2$v_3^{3/2} \Delta_{0.1}^{1/2} M_\odot$ for both SN 1954J and SN 1961V. As time passes, all these variability arguments become stronger since they have a minimum scaling that is quadratic in time. As noted earlier, fragmentation of the ejecta only accelerates the evolution. JWST observations at $\sim 20 \mu$m would also end any further speculation about dust. 

    Our DUSTY models favor relatively low temperatures of 5000 - 7000 K for both stars, so at least the observed stars cannot use a hot temperature to conceal a luminosity comparable to the progenitor. There is some room to have a hot companion, although not one as luminous as the progenitor. Particularly for very hot stars (Wolf-Rayet) stars, it also important to remember that stellar atmosphere models greatly underestimate the optical luminosities because they do not include reprocessing of the UV radiation by the stellar wind into optical emission and emission lines (see, e.g., \citeauthor{groh_fundamental_2013} \citeyear{groh_fundamental_2013}). It is marginally possible to have the hot companion suggested by \citet{humphreys_tale_2017} for SN 1954J. HST observations at shorter wavelengths would easily constrain these possibilities.
    
    The H$\alpha$ emission lines still seem a relatively compelling reason for the association of these stars with the transients. However, in the case of SN 1954J, \citet{humphreys_tale_2017} argue against the H$\alpha$ emission arising from ejecta produced in the transient. They instead require a hot companion star with a dense stellar wind (see section 4.4 of their paper). The apparent lack of evolution, at least in the case of SN 1954J, and the lack of any obvious source of of ionizing photons suggests that the line emitting material may not be associated with the observed transients. Instead, it could be material ejected in some earlier event that was then photoionized in 1954 or 1961. By placing the material further away, the evolution is slowed and it is easier to have long recombination times. For example, a solar mass of material at $R \sim 1/3$ pc in a thin shell ($\Delta = 0.1$) can produce roughly $L_{H\alpha} \simeq 10^{36}$ erg/s with a recombination time of about 200 years. At 1000 km/s, the shell would have been ejected $\sim$ 300 years ago. Shells, on smaller scales, quickly require a source of ionizing photons because the recombination times become shorter. We also found that the wind scenario considered by \citet{humphreys_tale_2017} also requires the present day stars to be much less luminous than the progenitors, independent of the dust properties.  
    
    This appears to leave only the options of survivors that are intrinsically less luminous than their progenitors, or that these stars are not the survivors. While we have not explicitly carried out the full calculations, it seems clear that we would reach the same conclusions for any of the historical progenitor candidates. Continued monitoring of both the broad band fluxes and the emission lines should steadily strengthen these conclusions. 

\section{Acknowledgments}
RAP would like to thank B. Shappee for his guidance in setting up the photometry and K.Z. Stanek for confirming the IR contamination of star 7. CSK is supported by National Science Foundation (NSF) Grants AST-1515876, AST-1515927, and AST-1814440. This work is based in part on observations made with the NASA/ESA \textit{Hubble Space Telescope} under program HST-GO-13477, and obtained from the Hubble Legacy Archive, which is a collaboration between the Space Telescope Science Institute (STScI/VASA), the Space Telescope European Coordinating Facility (ST-ECF/ESA), and the Canadian Astronomy Data Centre (CADC/NRC/CSA). This work is based in part on observations made with the Large Binocular Telescope. The LBT is an international collaboration among institutions in the United States, Italy, and Germany. The LBT Corporation partners are the University of Arizona on behalf of the Arizona university system,; the Istituto Nazionale di Astrofisica, Italy; the LBT Beteiligungsgesellschaft, Germany, representing the Max Planck Society, the Astrophysical Institute Potsdam, and Heidelberg University; the Ohio State University; and the Research Corporation, on behalf of the University of Notre Dame, University of Minnesota, and University of Virginia. 

\bibliographystyle{mnras}
\bibliography{late-time_obs.bib}

\bsp	
\label{lastpage}
\end{document}